\documentclass[twocolumn,showpacs,preprintnumbers,amsmath,amssymb,pra,aps,
    superscriptaddress,longbibliography, a4paper]{revtex4-2}
\usepackage[utf8]{inputenc}
\usepackage{physics}
\usepackage{dsfont}
\usepackage[unicode=true]{hyperref}
\hypersetup{colorlinks=true,linkcolor=blue,filecolor=blue,urlcolor=blue,citecolor=blue}
\usepackage[capitalize]{cleveref}
\usepackage{framed} 
\usepackage{natbib}
\usepackage{graphicx}
\usepackage{floatrow}
\usepackage[caption=false]{subfig}
\usepackage{tikz}
\usepackage{quantikz}
\usepackage{soul}
\graphicspath{{figs/}}

\newcommand{\CPHASE}{\texttt{CPHASE}}
\newcommand{\riSWAP}{$\sqrt{\texttt{iSWAP}}$}

\begin{document}
\title{Co-designing Transmon devices for control with simple pulses}
\author{Nicolas Wittler}
\affiliation{Theoretical Physics, Saarland University, 66123 Saarbr\"ucken, Germany}
\affiliation{Peter Gr\"unberg Institut -- Quantum Computing Analytics (PGI-12), Forschungszentrum J\"ulich, 52425 J\"ulich, Germany}
\email[Corresponding author: ]{n.wittler@pm.me}

\author{Shai Machnes}
\affiliation{Qruise GmbH, 66113, Saarbr\"ucken, Germany}

\author{Frank K. Wilhelm}
\affiliation{Theoretical Physics, Saarland University, 66123 Saarbr\"ucken, Germany}
\affiliation{Peter Gr\"unberg Institut -- Quantum Computing Analytics (PGI-12),
	Forschungszentrum J\"ulich, 52425 J\"ulich, Germany}

\date{October 14, 2024}

\begin{abstract}
In the current NISQ era, there is demand for functional quantum devices to
solve relevant computational problems, which motivates a utilitarian
perspective on device design: The goal is to create a device that is able to
run a given algorithm with state-of-the-art performance.
In this work, we use optimal control tools to derive the gate set required by
a toy algorithm and, in tandem, explore the model space of superconducting
quantum computer design, from dispersively coupled to stronger interacting
qubits, to maximize gate fidelity. We employ perfect entangler theory
to provide flexibility in the search for a two-qubit gate on a given platform
and to compare designs with different entangling mechanisms,
e.g., \CPHASE\ and \riSWAP. To ensure the applicability of our investigation,
we limit ourselves to "simple" (i.e., sparse parametrization) pulses and
quantify, where results differ from using the full complexity of piecewise
constant controls.
\end{abstract}

\maketitle

\section{Introduction}

When designing control layouts for superconducting quantum computing devices,
different approaches are being pursued.
Some designs, e.g., from IBM, are trying to achieve as much as is possible with
fixed-frequency qubits with microwave controls, avoiding the extra noise
introduced by flux lines. This has the challenge of no control over operating
sweet spots and couplings after fabrication. Adding flux lines incurs the cost
of increased decoherence to be able to move the working points of qubits around
at will. Following this design, adding additional tunable junctions to the chip
to act as couplers between data qubit junctions increases the footprint of the
chip, but promises to provide controlled interactions.  In addition, this
isolates the data qubits from flux noise \cite{krantzQuantumEngineersGuide2019}.

Creating devices that are well suited for a given task, as opposed to more general,
fundamental research, is desirable for the practical work on quantum algorithms. To
this end, tools are created to explore and fine-tune quantum device designs in
an integrated manner \cite{kunasaikaranFrameworkDesignRealization2023,
	niIntegratingQuantumProcessor2022, rajabzadehGeneralFrameworkGradientBased2024,
	guinnCoDesignedSuperconductingArchitecture2023}.

It is our goal to systematically analyze designs for quantum computing devices
that are fabricated for the purpose of executing quantum circuits or
algorithms. We will present our procedure with the examples of two common
design layouts: fixed qubits with fixed interaction (FQ), fixed qubits with a
tunable coupler (TC).
In this work, we'll consider the quantized Hamiltonian parameters -- Transmon
resonance frequencies, anharmonicities, coupling strengths -- as the search
space. Other efforts look to directly designing circuit quantities like
charging and junction energies \cite{menkeAutomatedDesignSuperconducting2021,
	rajabzadehGeneralFrameworkGradientBased2024} to produce devices with given
properties, e.g. certain types of many qubit interaction terms.

% Methods and entangler theory
Recent demonstrations have shown  that even in restrictive control settings,
almost all known gates are realizable \cite{weiNativeTwoqubitGates2023}. This
already presents a number of permutations of control schemes and platforms that
need to be evaluated. Providing a level playing field is a
non-trivial task, especially when trying to distill everything into a general
procedure.
Targeting a generalized perfect entangler can result in better outcomes than
the "textbook" two-qubit gate \cite{goerzOptimizingArbitraryPerfect2015,
	linLetEachQuantum2022, kairysEfficientQuantumGate2021}. When building a device
for a certain application, there's an advantage in considering the requirements
this poses to the gate-set \cite{googleaiquantumDemonstratingContinuousSet2020a,
	setiawanFastRobustGeometric2023}.

% Model search
This is more relevant when we add the model parameters to the search. In
\cite{goerzChartingCircuitQED2017a}, the authors coin the term "straddling"
regime -- the border between dispersive and strongly coupled -- and show that
this regime is optimal when employing full optimal control theory. We will
employ these methods to investigate the advantage of a tunable coupler setup
over fixed-frequency qubits, while limiting the control capabilities to simple
pulses with few parameters. We'll see what the effect of limiting the pulse
complexity has on the resulting model regime.

In \cref{sec:method}, we reproduce some of the theory for
creating entanglement. We present a general method to search for models that
facilitate both entangling and local gates. As an initial application, we apply
the method in \cref{sec:application} to a two-qubit
chip with and without an extra qubit to act as a tunable coupler.

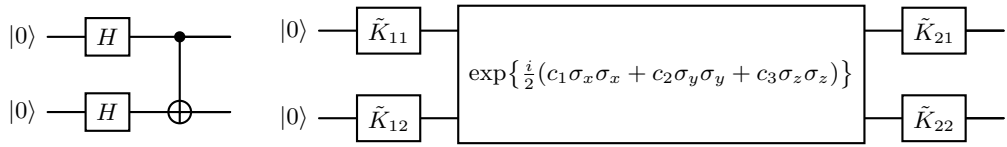
\begin{figure*}
	\begin{quantikz}
		\lstick{$\ket{0}$} & \gate{H} & \ctrl{1} & \qw \\
		\lstick{$\ket{0}$} & \gate{H} & \targ{} & \qw
	\end{quantikz}
	\quad%
	\begin{quantikz}
		\lstick{$\ket{0}$}
		& \gate{\tilde{K}_{11}}
		& \gate[wires=2]{\exp{\frac{i}{2}(c_1\sigma_x\sigma_x+c_2\sigma_y\sigma_y+c_3\sigma_z\sigma_z)}}
		& \gate{\tilde{K}_{21}}& \qw \\
		\lstick{$\ket{0}$} & \gate{\tilde{K}_{12}} & &\gate{\tilde{K}_{22}}&  \qw
	\end{quantikz}
	\caption{\textbf{Interpreting Lie group decomposition as circuit
			decomposition.} (a) Typical cell of a quantum computing circuit. (b)
		Decomposition with arbitrary entangling gate, given by the Weyl coordinates and
		modified local gates $\tilde{K}_{ij}$. Note that in a full algorithm circuit,
		layers of single-qubit gates can be compressed.}
	\label{fig:theory}
\end{figure*}

\section{Method}\label{sec:method}

For a functional quantum device, the fundamental tasks are:
(1) Define and optimize local gates,
(2) Define and optimize entangling gates,
(3) Find optimal system parameters to achieve both types of gates to high fidelity.

This presents a priori some conflicting interests: Entangling gates naturally
profit from strong coupling, as opposed to local gates, which benefit from
isolating subsystems. If the target is to run quantum circuits with high
accuracy, we'll need to find a compromise. Here, we will present a procedure to
achieve this, based on optimal control techniques.

The mechanism to generate entanglement for two-qubit gates generally depends on
the platform, so choosing a perfect entangler as the optimization target is
desirable to not limit possible solutions to a specific gate. For each setup,
there exists a procedure to derive entangling gates; the cross-resonance
(\texttt{CR}) gate for FQ and \CPHASE\ or \riSWAP\ for FQ and TQ, with AC or DC
controls \cite{dicarloDemonstrationTwoqubitAlgorithms2009,
	chowSimpleAllMicrowaveEntangling2011,
	ghoshHighfidelityControlled$ensuremathsigma^Z$Gate2013}.

\subsection{Perfect entangler theory}\label{sec:pe-theory}
A known approach is to consider an operator $U\in SU(4)$ (the computational
subspace, ignoring leakage levels for now) and characterize its non-local
properties by computing the Makhlin invariants $g_1, g_2$ and $g_3$
\cite{makhlinNonlocalPropertiesTwoQubit2002a}.

We are representing a general gate $U$ in the Bell basis as $U_B=Q^T U Q$, where
\begin{equation}
	Q = \frac{1}{\sqrt{2}}\mqty(
	1 & 0 &  0 &  i\\
	0 & i &  1 &  0\\
	0 & i & -1 &  0\\
	1 & 0 &  0 & -i
	) .
\end{equation}
In this basis, non-local gates are diagonal and local gates are orthogonal matrices.

The Makhlin invariants are then defined as
\begin{equation}
	\begin{aligned}
		g_1 & = \Re\frac{\Tr(U_B^T U_B)^2}{16\det(U)}                    \\
		g_2 & = \Im\frac{\Tr(U_B^T U_B)^2}{16\det(U)}                    \\
		g_3 & = \frac{\Tr(U_B^T U_B)^2-\Tr((U_B^T U_B)^2)}{4\det(U)} ~ .
	\end{aligned}
\end{equation}
These invariants characterize equivalence classes, e.g., \texttt{CNOT},
\CPHASE\ and \texttt{CR} share the same invariants, meaning they can be
transformed into each other with local gates.

We can generalize further by combining the invariants $\vec{g}=(g_1, g_2, g_3)$
into a single functional,
\begin{equation}\label{eq:pe-makhlin}
	d(\vec{g}) = g_3\sqrt{g_1^2+g_2^2}-g_1 ~,
\end{equation}
which vanishes for the invariants on the surface of the perfect entangler volume.
Inside the volume, the value has to be set to 0 manually
\cite{wattsOptimizingArbitraryPerfect2015}.

This functional is straightforward to compute for a given unitary and
represents a general measure for entangling performance. Some care must be
taken tough, depending on which side of the volume the current $\bar{U}$ lies.
We are interested in comparing the performance of single and two-qubit
operations during optimization and thus would prefer an equivalent fidelity of
the form $\abs{\Tr{V^\dagger \bar{U}}/ \dim{V}}^2$, where $V$ is the target
gate and $\bar{U}$ the time evolution of the controlled system.
One method is to use the perfect entangler functional for optimization and then
evaluate the final performance based on fidelity. Since we aim to find suitable
model parameters for both single-qubit and entangling gates at the same time,
we propose a direct numerical approach, the "tugboat" strategy.

\subsection{Tugboat optimization for perfect entanglers}
\label{sec:tugboat}
We write a general ideal target gate using the Cartan decomposition $V = K_1 A K_2$ with
\begin{equation}\label{eq:cartan}
	A = \exp{\frac{i}{2}(c_1\sigma_x\sigma_x+c_2\sigma_y\sigma_y+c_3\sigma_z\sigma_z)}
\end{equation}
where $K_{1,2}\in SU(2)\otimes SU(2)$. It can be shown, e.g., in \cite{zhangGeometricTheoryNonlocal2003a}, that the subalgebra $\{\sigma_x\sigma_x, \sigma_y\sigma_y, \sigma_z\sigma_z\}$ is sufficient to generate all of $SU(4)$ when combined with local rotations $K_{1,2}$. As an example, $\sigma_x\sigma_y$ can be realized from $\sigma_x\sigma_x$ by a single-qubit $\mathds{1}\sigma_z$.

\begin{figure}[b]
	\includegraphics[width=\textwidth]{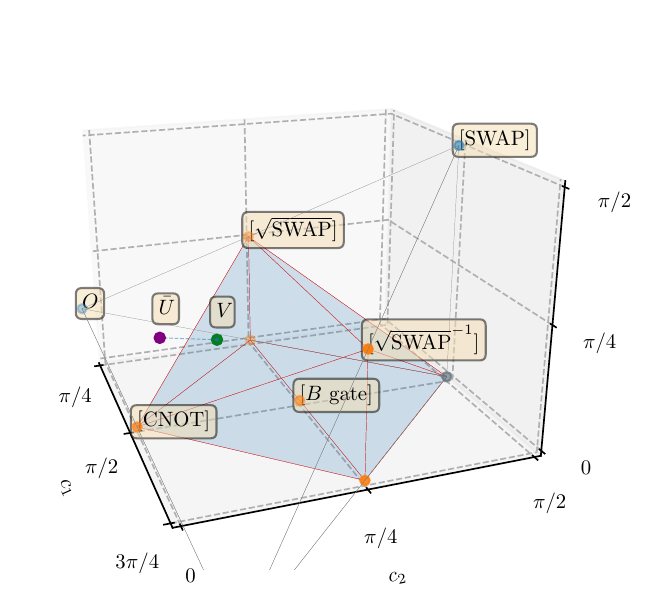}
	\caption[Weyl chamber of named gates to illustrate the optimization procedure.]{\textbf{Weyl chamber of named gates to illustrate the optimization procedure.} The labeled points indicate equivalence classes with respect to local rotations. The system dynamics $\bar{U}$ (projected into the computational space) are steered by the control and model parameters, while the target gate $V$ is varied over the shaded volume, bounded by \cref{eq:bounds}. This enables us to use $\abs{\Tr{V^\dagger(\vec{b},\vec{\gamma}) \bar{U}}}$ in a regular distance measure \cref{eq:2q-error}, indicated by the dashed line.
	}
	\label{fig:weyl}
\end{figure}

If we now confine the Weyl coordinates $c_i$ to the volume of perfect entanglers (shaded blue in \cref{fig:weyl}), we have a recipe to explicitly construct an arbitrary entangling gate. To this end, we propose the coordinate transformation
\begin{equation}
	\begin{aligned}
		c_1 & = \frac{b_1 + b_2}{2}            \\
		c_2 & = \frac{b_1 - b_2}{2}            \\
		c_3 & = b_3 (\pi/4 - \abs{c_2-\pi/4}))
	\end{aligned}
\end{equation}
to conveniently express the boundary conditions for the perfect entangler volume \cite{goerzOptimizingArbitraryPerfect2015} as
\begin{equation}\label{eq:bounds}
	\begin{aligned}
		\pi/4 \leq  b_1 & \leq \pi/2    \\
		0 \leq  b_2     & \leq \pi/4    \\
		0\leq b_3       & \leq \pi \- .
	\end{aligned}
\end{equation}

We will now consider a target gate, the "\emph{tugboat}", $V(\vec\gamma, \vec{b})$, where $\vec{b} = (b_i)$ are the transformed Weyl coordinates, confined to the volume of perfect entanglers, and local rotations $\vec{\gamma} = (\gamma^{(j)}_i)$ with $j=1,2$ enumerating the qubits and $i=x,y,z$ the rotation axis.

We assign an explicit fidelity error to a quantum gate $\bar{U}$ as
\begin{equation}\label{eq:2q-error}
	\epsilon_\text{II}(\bar{U}) = \min_{\vec{b},\vec{\gamma}}\qty{1-\abs{\Tr{V^\dagger(\vec{b},\vec{\gamma}) \bar{U}}/ \dim{V}}^2}
\end{equation}
where $\bar{U}$ is the time evolution of the system, projected into the computational subspace.
It is not resource intensive to compute a unitary matrix  $V$ in this representation, compared to the system time evolution $\bar{U}$ which requires several orders more computational power. We can thus easily find the closest perfect entangler representation to the implemented gate $\bar{U}$ by varying $\vec{b}$ and $\vec{\gamma}$ at each optimization step. This way we assign a regular fidelity measure to $\bar{U}$ while retaining the generality of the perfect entangler goal. An added benefit of this procedure it that we also explicitly extract the Weyl coordinates and local rotations of the entangling gate that allow us to identify the equivalence class.

\begin{figure*}[t]
	\includegraphics{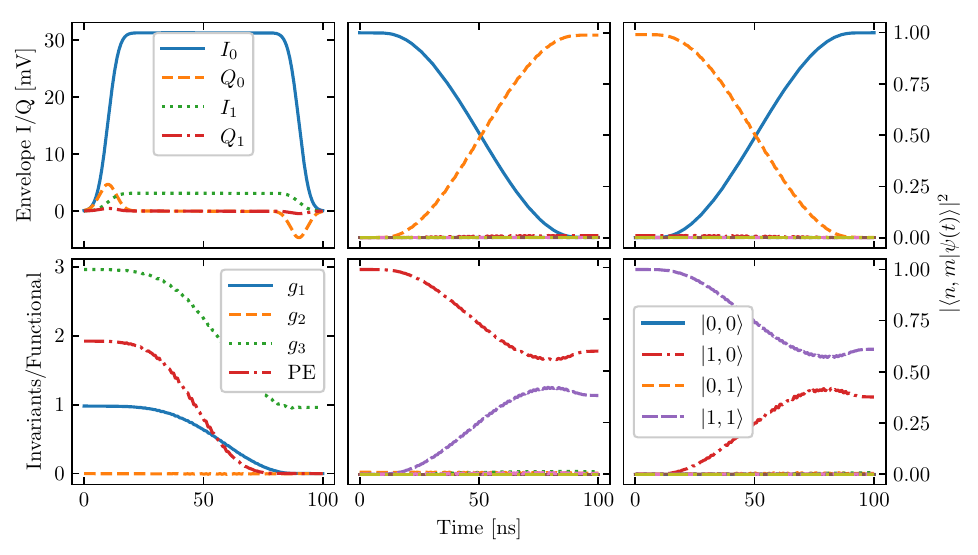}
	\caption{\textbf{Control fields and time-evolution of computational state populations of the two fixed Transmon device.} Both control fields are on resonance with the bare frequency of the second Transmon and the amplitude of the drive is increased by $(\omega_2-\omega_1) /g$. Their $I$ and $Q$ envelopes are shown on the top left.
		The panels on the right show the time evolution of populations, initialized in the four computational basis states.
		We observe a Rabi oscillation for Transmon 2 (right index, the \emph{target}), when Transmon 1 (left index, the \emph{control}) is in the 0 state.
		When the control Transmon is in the 1 state, only a partial population transfer occurs at a higher general Rabi frequency, realizing a general cross-resonance effect. To verify the entangling property, we plot the Makhlin invariants and the value of the perfect entangler (PE) functional from \cref{eq:pe-makhlin}.
	}
	\label{fig:dynamics}
\end{figure*}

\subsection{Optimal co-design}
In this formalism, we can now also explore the model parameter space by including it in the optimization. Assume the model is described by a Hamiltonian $H(\vec{\alpha}, \vec{\beta})$, with $\vec{\beta}$ being the vector of model parameters and $\vec{\alpha}$ the controls. The system dynamics are then given by
\begin{equation}
	U(\vec{\alpha}, \vec{\beta})
	= \mathcal{T} \exp{-\frac{i}{\hbar} \int dt H(\vec{\alpha}, \vec{\beta})}
\end{equation}
so formally the goal function depends on
$\epsilon_\text{II}\equiv \epsilon_\text{II}(\vec{\alpha}, \vec{\beta},\vec{b},\vec{\gamma})$.

Textbook representations of two-qubit gates, e.g. $\texttt{CNOT} =
	\ketbra{0}\mathds{1} + \ketbra{1}\sigma_x$, are implicitly written in the
interaction picture according to the Hamiltonian in the measurement basis,
denoted as $\tilde{H}$. We'll consider the readout basis to be the eigenbasis
of the undriven system.

We obtain the time evolution $\bar{U}= \mathcal{T}\exp(-i \int H dt)$ in the
product basis, since this basis will not change when we tune system parameters. 
Depending on the parameter
regime, writing the transformation that diagonalizes $H$ is not straightforward
enough to facilitate computing the gradient of the goal function with respect
to model properties.

To evaluate the optimization results for fixed parameters, we can simulate 
$\tilde H = S^\dagger H	S$, 
the diagonal (if applicable, dressed) Hamiltonian. We then write the goal
unitary as $ G = S \tilde G S^\dagger$ and, finally, the goal function as
\begin{equation}
	\varepsilon = 1 - \abs{\frac{1}{\dim G}\Tr{G^\dagger\bar{U}}}^2
\end{equation}

We can use $S\equiv S(\beta)$, the closed, exact form in
the computational subspace, that is dependent on $\omega_1, \omega_2$ and $g$.
\begin{equation}
	S = \mqty(
	\cos(b)& 0& 0& \sin(b)\\
	0 & \cos(a) & \sin(a) & 0\\
	0 & \-\sin(a) & \cos(a) & 0\\
	-\sin(b) & 0 & 0 & \cos(b)
	)
\end{equation}
with $a=\arctan(\frac{g}{\omega_2-\omega_1})$ rotating the exchange interaction
and $b=\arctan(\frac{g}{\omega_2+\omega_1})$ rotating the double excitations.
Since we performed the RWA on the interaction terms, here $b=0$.

As the fidelity of consecutive quantum gates is given by their product, the error for single-qubit gates ($\epsilon_\text{I}$) followed by an entangling gate (with error $\epsilon_\text{II}$) is
\begin{equation}
	\bar\epsilon
	= 1 - (1 - \epsilon_\text{I})(1 - \epsilon_\text{II})
	= \epsilon_\text{I}+\epsilon_\text{II} -\epsilon_\text{I}\epsilon_\text{II}
\end{equation}
where we'll neglect the product $\epsilon_\text{I}\epsilon_\text{II}$ as its contribution is small for errors 
below $1\%$ and use $\epsilon=\epsilon_\text{I}+\epsilon_\text{II}$ as our goal function.
\newpage
\section{Application}\label{sec:application}

We now apply the co-design procedure to explore designs for superconducting qubits. Model space consists of two Transmons with frequencies $\omega_{1,2}$, anharmonicities $\alpha_{1,2}$ and static interaction $g$ collected in a five-dimensional vector $\vec\beta = (\omega_1, \omega_2, \alpha_1, \alpha_2, g)$.
From the formal perspective, optimizing both model and controls at the same time constitutes a form of over-parametrization. For example, changes in drive tone and qubit frequency have the same effect on the goal function. Thus, we opt for simple controls, as opposed to a more general piecewise constant parametrization with hundreds of samples, so that
the dimensions of $\vec{\alpha}$ and $\vec{\beta}$ are in the same order of magnitude.

We will first run through the process for the two-qubit chip with fixed interaction strength, and then add a third qubit to act as a tunable coupler. To characterize the model by a single number, we take the reduced coupling
 $\chi = g_{12} / (\omega_2-\omega_1)$ and use the "straddling" regime $\chi=0.1$ as a divider between the
  dispersive ($\chi < 0.1$) and strong coupling ($\chi >  0.1$). When there is a tunable coupler, we use the effective interaction
   strength, where the coupler is adiabatically eliminated to have a comparable quantity
   \cite{mckayUniversalGateFixedFrequency2016a}.

\subsection{Local vs. entangling gates}

For universal quantum computing, we need to derive a set of local, single-qubit gates and an entangling gate. As a first application, we consider a device made up of two fixed-frequency Transmons with a static coupling. With all driving and coupling resonators eliminated, the Hamiltonian is $H =\sum_{j=0,1,2, g} H_j$ with
\begin{equation}\label{eq:transmon}
	\begin{aligned}
		H_0 & = \sum_i \omega_i a_i^\dagger a_i + \frac{\delta_i}{2}\qty( a_i^\dagger a_i - 1) a_i^\dagger a_i \\
		H_g & = g\qty(a_1^\dagger a_2 + a_1^\dagger a_2)                                                       \\
		H_i & = u_i(\vec\alpha, t)\qty(a_i^\dagger + a_i) \ ,
	\end{aligned}
\end{equation}
where we are just considering the exchange interaction.

The control field on qubit $i$ is 
$u(\vec\alpha, t) = A\qty[ I_i(t) \cos(\omega^d_i t + \phi_i) + Q_i(t) \sin(\omega^d_i t + \phi_i)]$ 
with a slow-varying, envelope $I_i$. As a basis function, we choose a flattop Gaussian shape
\begin{equation}
	I_i(t) = \erf\qty(-\frac{(t-t_\text{up})}{\sigma}) - \erf\qty(\frac{(t-t_\text{down})}{\sigma})
\end{equation}
with amplitude $A_i$ and ramps at $t_\text{up}$ and $t_\text{down}$ with a fixed width $\sigma=5$ ns. We also add an out-of-phase DRAG \cite{motzoiSimplePulsesElimination2009} correction as $Q_i = \lambda_i \partial_t I_i$.

First, we optimize both controls $\vec{\alpha} = (A_i, \omega_i, \phi_i,
	t_\text{up}^i, t_\text{down}^i, \lambda_i)$ and model parameters $\vec{\beta} =
	(\omega_i, \delta_i, g)$ to find single-qubit gates. We characterize the system
by its reduced coupling strength $g/(\omega_2-\omega_1)$.
To find the entangling gate, we employ the procedure described in
\cref{sec:tugboat}.
In \cref{fig:dynamics}
we show some
example dynamics. We initialize the target gate $V$ centered in the volume of
perfect entanglers at $\vec{c}/\pi=(1/2, 1/4, 0.05)$ and the local rotations
$\vec\gamma$ close to zero. After a set number of evaluations (here we found 4
to be a good value for convergence) in the gradient descent search, we update
the target unitary coordinates $\vec{c}$ and $\vec{\gamma}$ and refresh the
L-BFGS memory, if changing the target introduces significant error.

We want to include the effect of additional leakage in order to obtain
realistic control solutions. For each subsystem, we simulate three Transmon
levels labeled $\ket{0}$, $\ket{1}$ and $\ket{2}$. To ensure that system
dynamics stay within the embedded computational subspace, we write a running
cost
\begin{equation}
	L = \sum_k  \sum_{\lambda \in \Lambda} \abs{\braket{\lambda}{\psi(t_k)}}^2
\end{equation}
and add it to the goal function.
Population in the leakage levels $\Lambda = \qty{\ket{0, 2}, \ket{1, 2}, \ket{2, 0}, \ket{2, 1}}$ 
is evaluated at evenly spaced times $t_i$ during propagation.

\begin{figure}[t]
	\centering
	\includegraphics{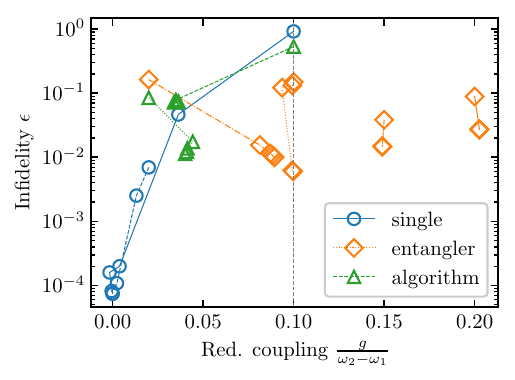}
	\caption{\textbf{Model space of a fixed frequency, two-qubit chip.} We
		optimize the parameters of two Transmons, their frequencies, anharmonicities
		and coupling, to realize local and non-local gates. Initializing the search
		around the straddling regime (dashed line), $\omega_2-\omega_1=500$ MHz $2\pi$
		and for different values of $g=50$ MHz $2\pi$, we observe that local gates
		trend towards the same convergence. Only trajectories that converge below an
		error of 0.1 are shown.}
	\label{fig:fq3waydispersive}
\end{figure}

In \cref{fig:fq3waydispersive}, we show the infidelities $\epsilon$ during optimization. 
Starting in a dispersive setting where $g/(\omega_2-\omega_1)<0.1$, we observe that
 optimizing for single-qubit gates trends towards isolated subsystems.
When we target a perfect entangler, we observe convergence to a higher
dispersivity, but still within the weak coupling regime. A combination of both
goal functions, representing a single cell
(\cref{fig:theory}) of some quantum computing
algorithm, results in a compromise between the two edge cases, both in terms of
dispersivity and reached fidelity. We also observe that the optimization
trajectory in the "algorithm" target seems to be a superposition of the two
layers. In this regime, the single-qubit gates can be realized to high
fidelity, so the entangling and "algorithm" errors are close.

When initializing the system in the "straddling" regime, i.e.
$g/(\omega_2-\omega_1)=0.1$ we find that we can recover the same convergence to
single-qubit gates. After a short exploration, the entangling gates return to
the same effective coupling value and perform slightly better than in the
dispersive initialization. The optimization for both however seems to get stuck
realizing a compromise. This regime has been shown to be optimal in the
presence of arbitrary control, but here, the chosen simple parametrization
might be not expressive enough to realize this potential.

\subsection{Flux-tunable coupler}

As seen in the previous section, choosing a fixed frequency, fixed coupling
design limits fidelities of algorithms, since the optimal parameter regimes for
single and two-qubit operations are different. The introduction of tunable
elements presents a solution, by allowing different working points for both
tasks. In principle, there are two designs: Adding a flux line to one of the
qubits to tune its resonance frequency or add a tunable coupler, another
non-data qubit with a tunable frequency.  However, the added flux lines also
provide a new avenue for noise to degrade coherence time of the qubits. To
compare to the fixed frequency setup in the previous section, we choose to
investigate a tunable coupler architecture and assume that the added noise
channel on the coupler qubit is reasonably isolated from the data qubits.

The flux-dependent frequency of a tunable element, is written as
\begin{equation}
	\omega_i(\Phi) = \qty(\omega_i^0 - \delta_i)\sqrt{\abs{\cos(\frac{\Phi}{\Phi_0}\pi)}}+\delta_i
\end{equation}
following the supplements of \cite{rolFastHighFidelityConditionalPhase2019},
where $\omega_i^0$ is the sweet spot frequency and $\Phi$ the external flux.
For simplicity, in this work we will directly consider the frequency offset
$\omega_i^c(t)$ as the control parameter, such that $\omega_i = \omega_i^0 +
	\omega_i^c(t)$.

As a characteristic quantity, we look at the effective interaction strength
\begin{equation}\label{eq:eff_interaction}
	J = \frac{g_1 g_2}{2}\qty(\frac{1}{\omega_1-\omega^0_\text{TC}}+\frac{1}{\omega_2-\omega^0_\text{TC}})
\end{equation}
induced between the two qubits \cite{mckayUniversalGateFixedFrequency2016a}.
Here, $g_1$ and $g_2$ being the coupling strength between qubit 1 and 2 and the
coupler, respectively.

\begin{figure}[t]
	\centering
	\includegraphics{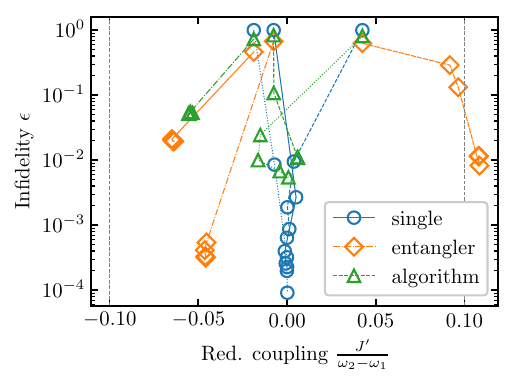}
	\caption{\textbf{Adding a tunable coupler to the model.} When we initialize
		the system in a sweet spot, where, according to Eq.
		\cref{eq:eff_interaction} the interaction between the
		qubits vanishes, good single-qubit gates can be reached. Entangling gates
		are trending towards effective couplings similar to the straddling regime.
		Optimizations starting in stronger coupling regimes has either stalled, or
		recovered to regions of weak coupling. Only trajectories that converge
		below an error of 0.1 are shown.}
	\label{fig:tc3waysweet}
\end{figure}

We put the idle point of the coupler $\omega^0_\text{TC}$ above the qubits
in \cref{eq:eff_interaction}, around the same difference in
frequency
\begin{equation}
	\omega_\text{TC}^0  - \omega_2  = \omega_2 - \omega_1  ~ .
\end{equation}
If, in addition, there's a static coupling $g_{12}$ between the qubits, the
total effective interaction is
\begin{equation}
	J' =g_{12} + \frac{g_1 g_2}{2}\qty(\frac{1}{\omega_1-\omega^0_\text{TC}}+\frac{1}{\omega_2-\omega^0_\text{TC}})
\end{equation}

For $g_1$ and $g_2$, we pick similar values to the coupling constant in the
previous example, and we select the residual interaction $g_{12}$ to initialize
the optimization in different $J'$ regimes. To entangle the qubits, we modulate
$\omega_c$ with an AC tone close to their difference frequency
\cite{mckayUniversalGateFixedFrequency2016a}, shaped by the same flattop
envelopes as before. Again, we can realize single and entangling gates to good
fidelities.
Single-qubit gates can reach good fidelities here, even exploring the region
around the sweet spot and returning. Once the interaction strength is high
enough, we can also entangle the qubits with a higher fidelity than the case
without the tunable coupler. For the "algorithm" optimization, the same
competition between local and non-local operations results in a stagnation at
sub-par fidelity.
The trajectory of the combined case is again alternating between the individual
trajectories, see \cref{fig:tc3waysweet}.

In both cases, we see that the approaching the strong coupling regime presents
problems in realizing single-qubit gates, at least with the simple pulse shapes
we explored here.

\section{Summary and Outlook}
We have shown a general procedure to investigate model design space for an
application like superconducting qubits, where the properties of human-made
devices can be targeted in fabrication. Instead of a full optimal control
approach, we investigate if previous results about preferable regimes hold for
simple pulse shapes. Compared to previous work, we choose a direct search in
the Weyl space to find the closest perfect entangler to the gate our test
system can produce. This way, we avoid some mathematical difficulties, since
the expression of the functional \cref{eq:pe-makhlin} is only
defined outside the volume. Inside the volume, the value is 1 by definition.
The direct use of \cref{eq:2q-error} does not require these conditional expressions.

The application to the two-qubit model, optionally adding a tunable coupler,
shows reliably that good single-qubit gates can be reached when the effective
interaction is turned almost off. This is as expected but also holds, when
initializing a search some distance away from this regime. By the same
reasoning, targeting entangling gates trends towards a similar point -- the
previously identified "straddling" regime from different initial values. There,
we notice the necessity to restart optimizations, in the case of early
termination. The use of simple pulses seems to limit the exploration of
parameter space and prevents a good regime for the operation of both local and
non-local quantum gates. To make a more systematic claim, a wider range of
cases -- control schemes and device designs -- needs to be tested. For example,
several schemes exist to mitigate the effect of spurious $ZZ$ interactions to
allow single-qubit control on more strongly interacting systems
\cite{mundadaSuppressionQubitCrosstalk2019,
	kandalaDemonstrationHighFidelityCnot2021}.

Even limited to the superconducting platform, several design paradigms to
realize quantum computing devices exist. The Transmon, with its specific
$E_C/E_J$ ratio, is only one possible design. It is the qubit design most
demonstrations of applied quantum computing use. There are other promising
paradigms, such as the Fluxonium, which combines insensitivity to external
noise with a large anharmonicity \cite{baoFluxoniumAlternativeQubit2022}. So, a
different search space can be explored with the method presented here by
directly optimizing Josephson junctions, capacitances and inductances.

\section{Acknowledgements}
We thank our colleagues, particularly A. Mishra, for discussion and comments on
the manuscript.

This work is supported by funding from the German Federal Ministry of Education
and Research via the funding program quantum technologies --
from basic research to the market -- under contract numbers 13N15680 ”GeQCoS”.

\end{document}